# Can Machine Learning Catch Economic Recessions Using Economic and Market Sentiments?

Kian Tehranian[1]

## 1. Abstract

Quantitative models are an important decision-making factor for policy makers and investors. Predicting an economic recession with high accuracy and reliability would be very beneficial for the society. This paper assesses machine learning technics to predict economic recessions in United States using market sentiment and economic indicators (seventy-five explanatory variables) from Jan 1986 – June 2022 on a monthly basis frequency. In order to solve the issue of missing time-series data points, Autoregressive Integrated Moving Average (ARIMA) method used to backcast explanatory variables. Analysis started with reduction in high dimensional dataset to only most important characters using Boruta algorithm, correlation matrix and solving multicollinearity issue. Afterwards, built various cross-validated models, both probability regression methods and machine learning technics, to predict recession binary outcome. The methods considered are Probit, Logit, Elastic Net, Random Forest, Gradient Boosting, and Neural Network. Lastly, discussed different model's performance based on confusion matrix, accuracy and F1score with potential reasons for their weakness and robustness.

✉   Kian Tehranian, tehraniank@g.ucla.edu

[1]   Department of Economics, University of California – Los Angeles (UCLA), Los Angeles, CA, United States

**Table Of Contents**

## 2. Introduction

According to Nyman and Ormerod (2016), it is around fifty years we are living in a data driven economy. Economists usually disagree about how the economy operates and different economic policies, some believe in Keynesian model, some in neoclassical economic model. However, economists try to enhance the accuracy of models, but not a particular method has a better accuracy than others in all times.

After the recent economic instability and increasing money supply in the US economy, the issue of forecasting recessions has major implications in terms of economic policy as well as economic stability. Notifying upcoming economic downturn can be very effective to policymakers and investors as they can take the necessary actions either in the form of monetary and fiscal policy or hedging against a probable outcome. For this reason, many researchers have tried to exploit the information from market and economic sentiments to forecast economic downturns.

Many studies focus on specific market variables as explanatory variables to predict recessions using machine learning algorithms. However, in this paper I explored seventy-five market and economic indicators such as economic variables of major US trade partners, recession resilient good's prices, commodities, stocks, equity indexes, fixed income indexes, and currencies from Jan 1986 to June 2022 June.

A particular problem is the lack of data availability with high frequency. On one hand we started collecting economic and market data from fifty years ago and the economic variables have low frequency, on the other hand, as machine learning classification algorithms mostly use decision tree base approach, we would require a large dataset to be able to train a robust and reliable model.

The main contribution of this paper is to train models based on different machine learning algorithms and examine whether the previously mentioned indicators can improve forecasting accuracy for the binary outcome of economic recessions.

Also, as a few variables were not available since the beginning of our dataset (Jan 1986), I used Autoregressive Integrated Moving Average (ARIMA) model to backcast time-series variables with respect to trend, seasonality, and cycles.

In the next sections I will explain and discuss about each variable, models, and results in details.

## 3. Literature Review

There has been always debate about the universal definition of recession. The idea of recession as two negative quarters of GDP growth emerged and became popular as a result of Julius Shiskin's article in 1974 that looked at a list of factors to capture a recession. Donoghue (2009) argued the 'two quarter negative growth rate' of GDP is not appropriate and causes problems. She believed the definition of recession fails to take into account factors other than GDP, also, the definition does not recognize 2001 recession in America and 1970s severe recession in Japan. She suggested National Bureau of Economic Research (NBER) approach seems better, but it wouldn't inform us until we are in a recession, and sometimes not until it is already over. According to Layton and Banerji (2003), recession is a philosophical notion that characterizes an economy rather than an output measure.

Ormerod and Mounfield (2000) indicated, the GDP growth is dominated by noise rather than signal. However, Nyman and Ormerod (2016) illustrated although there is upper limit in prediction accuracy, non-linear machine learning methods are useful in providing early warnings of recession to both policy makers and market participants. They used publicly available data consist of market sentiments and economics variables to predict GDP growth. They suggested between three and six quarters ahead although the correlation between actual and prediction are low, they are very different from zero; also, whenever they predicted a recession, one did occur. Their algorithm predicted a large economic downturn would occur in the first half of 2009 six quarters ahead in late 2007. Estrella and Mishkin (1996, 1998) used macroeconomic indicators with Probit method to forecast recession and they found out some well-known macroeconomic indicators and stock prices are good indicators in one to three quarters time horizon, however, for the longer time horizons the slope of yield curve appeared to be the major recession indicator. Additionally, Fornani and Lemke (2010) extended probit approach by developing regressors using VAR. They realized adding lags of treasury term spreads improves the predictability of recessions at shorter time horizons, while the treasury term spread has the highest predictive power in four to six quarters ahead time horizon. Puglia and Tucker (2020) by using treasury yield curve spreads (10yr treasury vs 3m treasury), S&P 500 index return values, excess bond premium, and federal fund rates, argued machine learning and neural-network methods due to their flexibility can capture non-linear nature of important features in macroeconomic data, but probit method cannot. On the other side, they noted considering the monthly basis frequency of economic variables, the algorithms are designed for large dataset applications and even then, they may over-fit data. Besides, worth noting all methods used in their research signaled rise of recession probability as the term spreads flatten, meaning financial conditions tighten or macroeconomic factors deteriorate.

Coulombe, Marcellino and Stevanovic (2021) studied Covid recession using machine learning on 112 macroeconomics and financial indicators in nine categories, nonlinear methods with the ability to extrapolate had better performance, unlike linear methods. In addition, Malladi (2022) across 134 variables referred as the 'Stock-Watson dataset' (Stock & Watson, 2006) detected labor market (average weekly hours in manufacturing, housing starts in the west) is the top Covid recession indicator and 107 variables didn't have much prediction power. Plus, based on his best performed machine learning algorithm, linear support vector machine (LSVM), a recession was predicted six months before the official NBER announcement, also the algorithms warned S&P 500 severe downturn two months before it happened. Worth noting, he pointed out predicting a recession is approximately three times harder than predicting a stock market crash based on the false discovery rate (FDR) or a type 1 error.

## 4. Data

There is a debate regarding what should be considered as a recession. Generally, recession has been a period of two quarters of negative GDP growth. However, this definition is no longer popular and many economists such as Donoghue (2009) challenged the idea as it failed to recognize 2001 recession. Since then, some suggested recession would be recognized when there are several consecutive slowing growth periods. These days a widely used alternative option that is used in this paper is Business Cycle Dating Committee at the National Bureau of Economic Research (NBER). It uses a broader definition of 'a decline in economic activity that lasts more than a few months and visible in real GDP growth, real personal income, employment, industrial production, and wholesale-retail sales. As a result of NBER recession indicator, United States had four recessions (forty periods) since Jan 1986 to June 2022.

Due to data availability constraint and monthly frequency of most economic indicators, the dataset is considered on a monthly basis from Jan 1986 to June 2022 (438 periods). Also, in order to capture different dimension of an economic recession I examined seventy-five variables consist of changes in US and its major trade partner's economic variables, recession proof goods price returns, base commodity price returns, fixed income market indicators, equity index returns, currency pair returns between US and its major trade partners, and lastly recession proof stock returns. Some of data is publicly available at Federal Reserve Economic Data (FRED) website, while others are taken from Bloomberg Terminal.

**4.1 Variable Explanation**

Economic Indicators: In order to illustrate a full picture of US economic characteristics and with an inspiration of 'Stock-Watson dataset' (Stock & Watson, 2006), I investigated Consumer Price Index YoY change (CPI), Personal Consumption Expenditure YoY change (PCE), Industrial Production YoY change, Non-Farm Payroll YoY change, Unemployment Rate %, Personal Income YoY change, Personal Savings YoY change, Net Trade in Goods $B (also called merchandise trade), Net Trade in Goods YoY change, Uni Michigan: Consumer Sentiment YoY change, Monthly Supply of New Houses YoY change, New Housing Unit Started YoY change, Existing One Family Home Sales YoY change.

Major trade partner's economic indicators: According to Observatory of Economic Complexity (OEC), largest United States trade partners (export and import) are Canada, Mexico, China, Japan, Germany, and United Kingdom, the change in these economies would affect US economy. Therefore, I considered CPI change, Unemployment Rate %, and Industrial Production YoY change in these economies except for China due to lack of data availability.

Goods Prices: As Ahrens (2007) illustrated, empirically there are some tradable items with high demand during recessions that can be considered as recession-proof goods such as Food Price YoY change, Alcoholic Beverages Price YoY change, Tobacco and Smoking Products Price YoY change, Cosmetics Price YoY change, Cocoa Futures monthly return, and Coffee Arabica Futures monthly return.

Commodity Prices: United States as a major industrial economy is considered as one of largest importer and exporter of commodities, and therefore the price changes would largely affect the capital-intensive industries. As a true representative of commodity market some base metal and agricultural futures contract prices considered. In terms of agricultural products since US economy is major producer of Soybean, Corn, and Lumber, their monthly returns are considered. Besides, in consideration of base metals futures contract's monthly returns of Gold, Silver, WTI Crude Oil, LME Copper, LME Aluminum, LME Nickel are included. However, in the case of LME Copper, LME Aluminum, and LME Nickel, prices were not available for four, eighteen, and thirteen periods respectively from the beginning of the analysis period which my solution was to backcast with ARIMA time-series model that takes into account trend, seasonality, and cycles (the model prediction details are explained later).

Fixed Income Market Indicators: Since yield curve is reflecting a wide range of information about future market expectations, inflation expectations and broad financial system conditions, 3m, 2yr, 5yr, 10yr, 30yr treasury government bonds are considered. Also, Bloomberg 10yr municipal index YTW and the ratio of municipal index over 10yr treasury bond (10yr Muni/10yr TSY) are used as municipal market representative and a widely used ratio for the richness of the municipal yield versus risk free government treasury note. Lastly, 3months federal fund rate (FFR) and treasury term differences that are empirically used as indicators of recession probability and financial conditions (Puglia and Tucker, 2020) are also considered (2yr-3m TSY, 10yr-3m TSY, 10yr-2yr TSY, 30yr-2yr TSY, FFD-3m TSY).

Currencies: five major US trade partner's currency pair monthly returns, plus dollar index monthly changes are considered to reflect the change in the situation between economies (EURUSD return, USDJPY return, GBPUSD return, USDCAD return, USDMXN return, DXY return).

Equity Market Indicators: US and its largest trade partner's well known equity index monthly returns are chosen (SPX Index return, UKX Index return, TPX Index return, DAX index return, and HIS Index return). In addition, SPX price to earnings ratio (P/E ratio) is used to reflect high/low multiple valuations in the US market, and also, according to empirical studies some industries generally do well during recession periods that I used some well-known large company names as recession proof equities. The stocks include United Health (UNH US Equity), Home Depot (HD US Equity), McDonald's (MCD US Equity), Johnson & Johnson (JNJ US Equity), Procter & Gamble (PG US Equity), Walmart (WMT US Equity), Coca-Cola (KO US Equity), JPMorgan Chase & Co (JPM US Equity), FedEx (FDX US Equity), IBM (IBM US Equity), Microsoft (MSFT US Equity), and Apple (AAPL US Equity). However, there was a missing data point for MSFT US Equity due to the company IPO on March 1986, and so to overcome the three period missing data I again used ARIMA model to backcast the time-series data with respect to its trend, seasonality, and cycles (explain the model selection later).

**4.2 Dealing with missing data - Backcasting**

First step in dealing with missing a few data points on four variables was to look at distribution and normality of variables. In the case of a normally distributed variable, it is common to replace mean or median, as used by Coulombe, Marcellino and Stevanovic (2021). However, based on the Jarque-Bera normality test results (Table 1) and distribution plots (Fig 1) all four variables are non-normal (p-values below 5%).

*Table 1*: Jarque-Bera test result

| LME Copper: | P-value: 0.0 | LME Aluminum: | P-value: 0.0072 | LME Nickel: | P-value: 0.0 | MSFT US Equity: | P-value: 0.0 |
|---|---|---|---|---|---|---|---|
| Mean: 0.65 | Std: 7.06 | Mean: 0.27 | Std: 5.82 | Mean: 1.12 | Std: 11.52 | Mean: 2.27 | Std: 9.6 |
| Max: 35.48 | Min: -35.55 | Max: 19.46 | Min: -21.8 | Max: 101.34 | Min: -27.26 | Max: 51.58 | Min: -34.35 |

*Fig. 1* distribution plot per variable

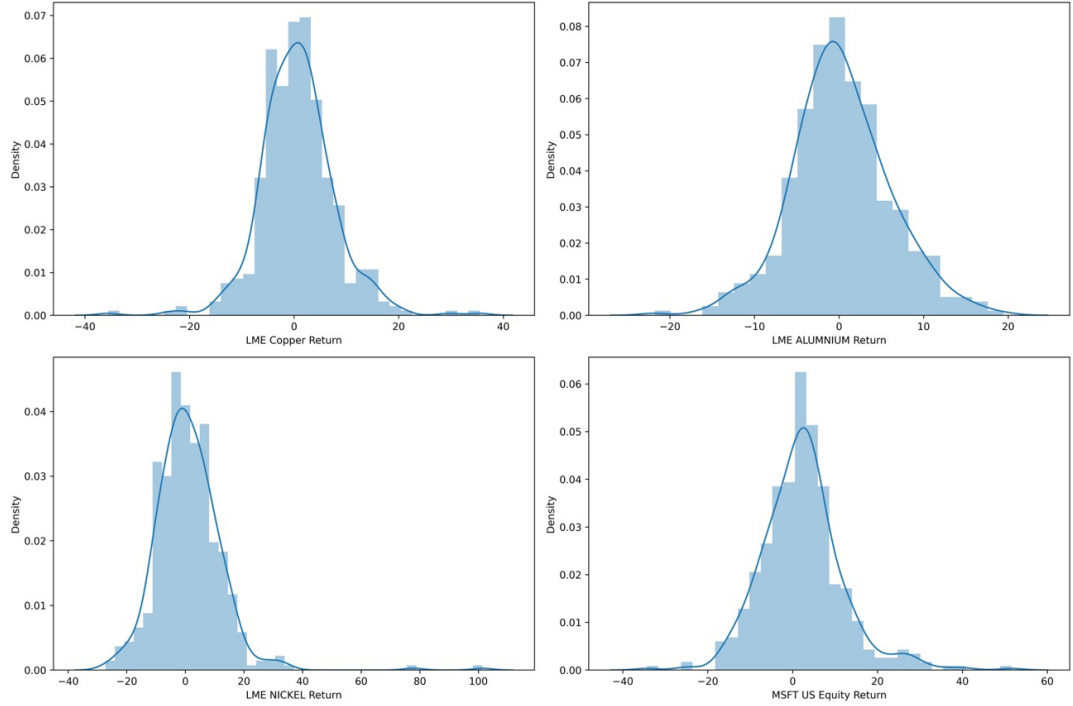

Therefore, since variables are time-series, Autoregressive Integrated Moving Average method (ARIMA) is used to train a model with respect to trend, seasonality, and residual components and then backcasting the missing values.

ARIMA(*k, d, q*) is produced by adding differential method to ARMA(*k, q*) (short for autoregressive moving average), which is combination of AR(k) and MA(q). ARMA model assumes $X_t$ is resulted from noisy linear combination of last *k* observations. The ARMA model is generated via from formula below:

$$X_t = \sum_{i=1}^{q} \beta_i \epsilon_{t-i} + \sum_{i=1}^{k} \alpha_i X_{t-i} + \epsilon_t$$

(1)

Where $\epsilon_t$ are noise term, and $\beta_i$, $\alpha_i$ are coefficients of MA(q) and AR(k) models respectively.

ARIMA(k, d, q) model is the adding differential method to ARMA(k, q) method. For example, first order difference of $X_t$ can be calculated as $\nabla X_t = X_t - X_{t-1}$ and the second order difference as $\nabla^2 X_t = \nabla X_t - \nabla X_{t-1}$. Therefore, the model ARIMA (k, d, q) can be written as:

$$X_t = \nabla^d X_t + \sum_{i=0}^{d-1} \nabla^i X_{t-1}$$

(2)

Where $d$ is the d-th order differential of observation and $\nabla^d X_t$ is:

$$\nabla^d X_t = \sum_{i=1}^{q} \beta_i \epsilon_{t-i} + \sum_{i=1}^{k} \alpha_i \nabla^d X_{t-i} + \epsilon_t$$

(3)

In this paper, Auto-Arima package is used which runs Augmented Dickey-Fuller test (ADF test) to determine stationarity of data and then finds optimal value of 'd' (number of nonseasonal differences). Also, it chooses the optimal values for 'k' and 'q' (number of autoregressive terms and number of lagged forecast errors respectively) with the goal to find the lowest AIC and BIC values, which are representative of goodness of the fit. After deciding on ARIMA (k,d,q) parameters, models are trained and backcasted for each four variables, the summary tables and plots are shown in the Table 2 and Fig 2 respectively.

*Table 2: ARIMA model summary table for each variable*

|  | LME Copper Return | LME ALUMNIUM Return | LME Nickel Reurn | MSFT US Equity Return |
|---|---|---|---|---|
| Observations: | 434 | 420 | 425 | 435 |
| Model: | ARIMA(1,0,1) | ARIMA(1,0,1) | ARIMA(1,0,1) | ARIMA (0,0,1) |
| Log Likelihood: | -1486.16 | -1357 | -1632.18 | -1567.55 |
| AIC: | 2982.33 | 2724.01 | 3274.36 | 3145.11 |
| BIC: | 3002.55 | 2744.06 | 3294.47 | 3165.35 |

*Fig. 2 ARIMA model predicted value plot for each variable*

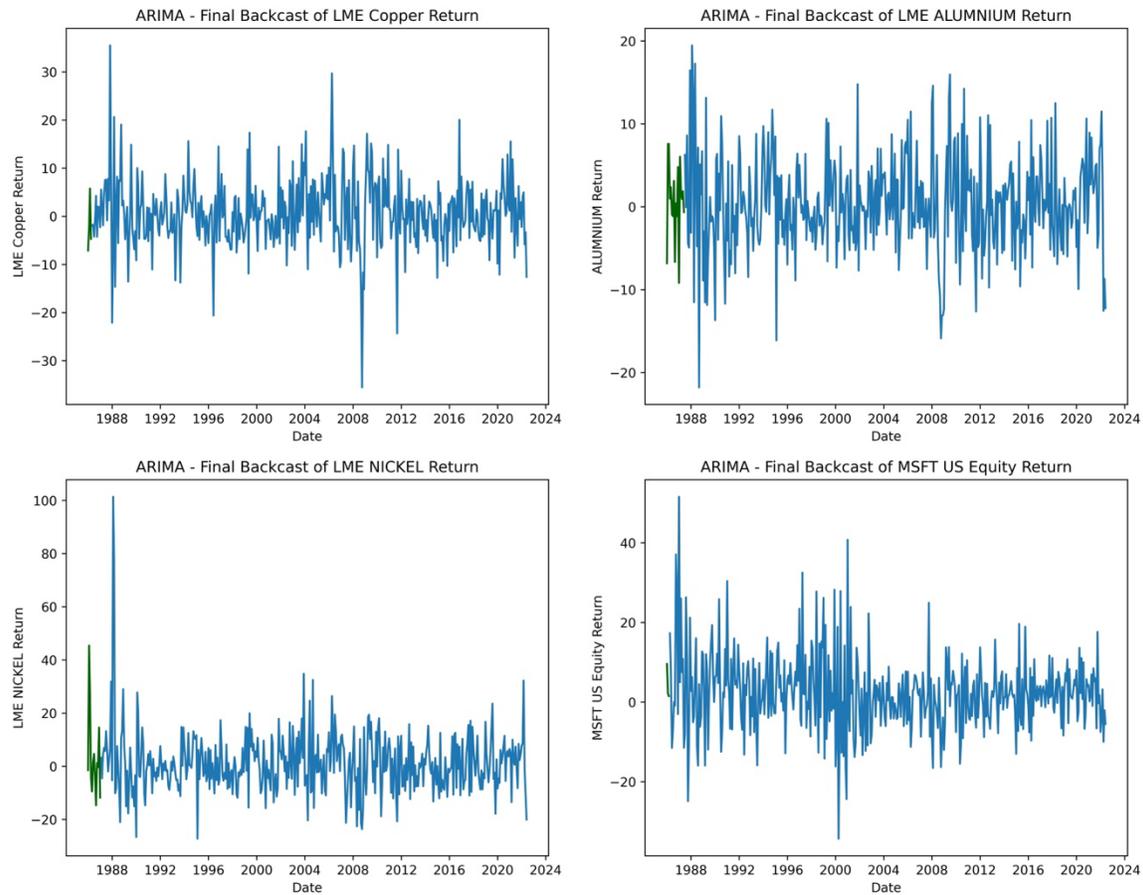

The backcasted values are replaced with the missing points in the original dataset. As a result, the dataset is fully prepared and ready to start analysis. In the next section I will discuss how I chose which variables to be used in the models, afterwards, I will discuss about results and variable importance from different methods.

## 5. Methodology and Result

**5.1 Variable Selection**

The type of dataset I am dealing with is high-dimensional data. In order to extract useful information, first statistical techniques should be used to reduce the noise and redundant variables. By using only important variables we can usually improve our model precision and decrease the complexity and so easier to interpret.

First start by looking at correlation matrix of two more obvious duplicate columns, Personal Consumption Expenditure YoY (PCE) versus Consumer Price index YoY (CPI) and Net Trade in Goods $B versus Net Trade in Goods YoY, and choose the ones that are more relevant to our response variable (Fig 3). PCE and CPI using different way to measure inflation, also, between the gross net trade amount and the year over year change, the more relevant one to the response variable should be considered.

*Fig. 3* correlation matrix of response variable (Recession) and four explanatory variables

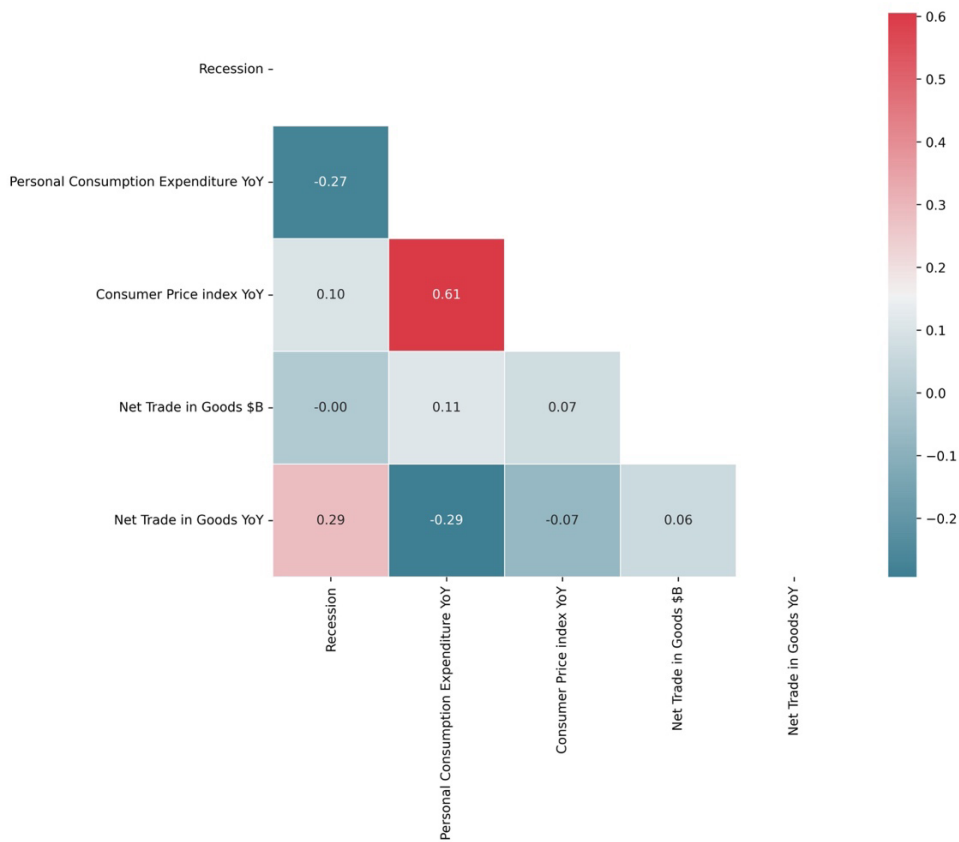

Based on the correlation magnitude between response variable (Recession) and explanatory variable pairs, Net Trade in Goods YoY and Personal Consumption Expenditure YoY are chosen.

Generally, in order to reduce the dimensionality of the data Principal Component Analysis or Singular Value decomposition technics are used; however, these methods are unsupervised feature selection that do not take into account the information between explanatory variables and response variable. In this paper, I used Boruta algorithm. Boruta algorithm is a random forest classification algorithm that captures features with respect to response variable (Liaw and Wiener, 2002). It measures attribute importance by adding randomness to the system and collecting the result with the goal to reduce the random fluctuations and correlations. The algorithm run random forest classifier and gather Z-score of the shuffled copies of the features and the original features, then finds the maximum Z-score among shadow attributes (MZSA), afterwards for undetermined importance variables it performs a two-sided test of equality with the MZSA and deem the attributes with higher importance compared to MZSA as important variables. Lastly the algorithm repeats this process until importance is assigned to all attributes.

After running it on the dataset, fifteen features are chosen which then I only work with these variables as the important ones. Also, correlation between these fifteen selected features is checked to refrain from multicollinearity (Fig 4).

*Fig. 4 correlation matrix for fifteen variables selected by Boruta*

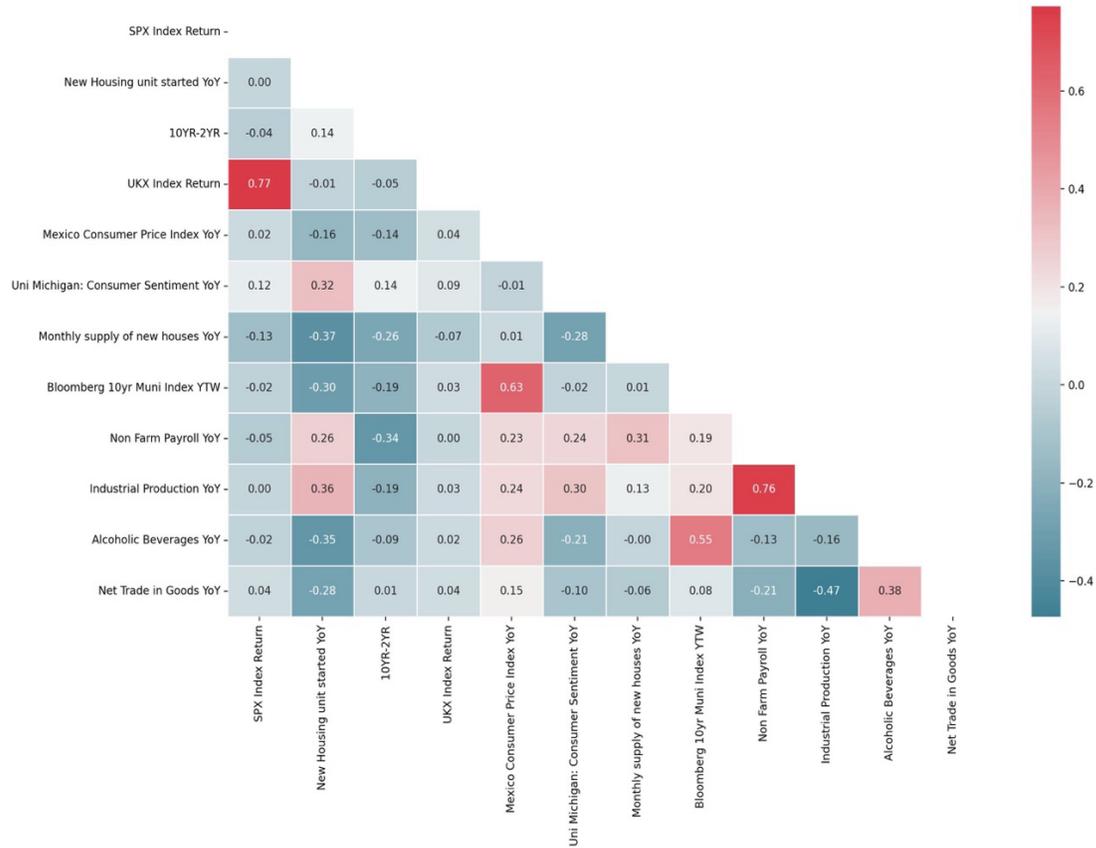

There are three variables (USGG10YR Index, USGG30YR Index, Canada Industry Production Excluding Construction YoY) that have high correlation (above 0.8) with other variables. Therefore, after excluding these three variables, ended up with only twelve most relevant variables to start training various models (Table 3).

*Table 3: important features selected by Boruta algorithm and correlation matrix*

| Selected Features | | | |
|---|---|---|---|
| SPX Index Return | Non-Farm Payroll YoY | Mexico Consumer Price Index YoY | Uni Michigan: Consumer Sentiment YoY |
| Bloomberg 10yr Muni Index | New Housing unit started YoY | Monthly supply of new houses YoY | 10YR-2YR |
| UKX Index Return | Industrial Production YoY | Alcoholic Beverages YoY | Net Trade in Goods YoY |

## 5.2 Models

At this point the dataset is fully ready for training models. As the relationship between explanatory and response variables are unknown, I tried probability regression base, machine learning classification and neural network models. In terms of probability regression base models, Probit and Logit models are trained, also, elastic net, random forest and gradient boosting models are tested.

I used two different result reporting metrics called accuracy and F1 score, that both uses confusion matrix to assess the prediction:

|  |  | Actual | |
|---|---|---|---|
|  |  | Positive | Negative |
| Prediction | Positive | TP | FP |
|  | Negative | FN | TN |

Since the outcome is binary the accuracy metric compares the sum of true positive and true negative recession predictions with the total number of predictions:

$$Accuracy = \frac{TP + TN}{TP + FP + FN + TN}$$

(4)

F1 Score is usually used for uneven variable distribution that uses two ratios to report a more reliable score:

$$F1\ Score = 2 \times \frac{Recall \times Precision}{Recall + Precision}$$

(5)

Where *Recall* and *Precision* are as follow:

$$Recall = \frac{TP}{TP + FN}$$

(6)

$$Precision = \frac{TP}{TP + FP}$$

(7)

**Probit & Logit:** The regression methods used are Probit and Logit which have a likelihood (probability) outcome for the response variable (1 as recession, 0 as non-recession). As Probit and Logit are fully explained by Kovacova and Kliestik (2017), the probability function is defined as below:

$$P_1 = \frac{e^z}{e^z + 1} \; , \; P_0 = 1 - P_1$$

(8)

$$Z = \beta_0 + \beta_1 x_1 + \beta_2 x_2 + \cdots + \beta_n x_n$$

(9)

Where $\beta$s are the coefficients of the regression in the way to maximize the likelihood. I considered a likelihood of more than 0.5 to be considered as recession and below 0.5 as no recession.

Before running the models, in order to check for linear model misspecification, Reset-Ramsey test is used. It helps to examine if the linear combination of explanatory variables is mis-specified.

Reset-Ramsey test (Sapra ,2005) is comparing the regression function with higher order version of the same regression function, then illustrate if the model is mis-specified by rejecting a null hypothesis of $H_0: \gamma_1 = 0$ versus $H_1: \gamma_1 \neq 0$ in the following formula:

$$\eta_{1i} = \beta' x_i$$
$$\eta_{2i} = \beta' x_i + \gamma_1 \eta_{1i}^2$$
$$\eta_{3i} = \beta' x_i + \gamma_1 \eta_{1i}^2 + \gamma_2 \eta_{1i}^3$$

(10)

After running the test, the results are stated below (Table 4):

*Table 4: Reset-Ramsey test results for different regression powers*

| Reset-Ramsey Test | |
|---|---|
| Power 2 | P-value: 2.69e-29 |
| Power 3 | P-value: 2.42e-32 |
| Power 5 | P-value: 2.66e-51 |

According to Reset-Ramsey test the linear model is correctly specified because not only the null hypothesis of power two is rejected (p-value below 0.05), but also as the power goes up the p-value decreases. So, I conclude the linear combination of variables are not mis-specified.

Now, I split the train and test set to 75% and 25% respectively and run Probit model with linear variable combination. Probit model analyzes the relationship between binomial response variable and explanatory variables. Considering a log likelihood outcome of dependent variable, the formula (Kovacova and Kliestik, 2017) is given by:

$$P_1 = 1 - \phi(-x, \beta) = \phi(\beta_0 + \beta_1 x_1 + \beta_2 x_2 + \cdots + \beta_n x_n)$$

(11)

$$\phi(x, \beta) = \int_{-\infty}^{x,\beta} \frac{1}{\sqrt{2\pi}} e^{-\frac{x^{-2}}{2}} dx$$

(12)

The summary table and model evaluation metrics are as below (Fig 5).

*Fig. 5 Probit model summary table, out of sample confusion matrix, prediction accuracy, and F1 score*

```
results_probit.summary():
Probit Regression Results
===============================================================================
Dep. Variable:                Recession   No. Observations:                  329
Model:                           Probit   Df Residuals:                      316
Method:                             MLE   Df Model:                           12
Date:                  Mon, 29 May 2023   Pseudo R-squ.:                  0.7456
Time:                         22:26:33    Log-Likelihood:                -23.756
converged:                         True   LL-Null:                       -93.366
Covariance Type:              nonrobust   LLR p-value:                 8.605e-24
===============================================================================
        coef    std err          z      P>|z|      [0.025      0.975]
```

```
------------------------------------------------------------------------------------------
Intercept                              -4.1213      1.109    -3.716    0.000    -6.295    -1.948
UKX_Index_Return                       -0.1546      0.090    -1.711    0.087    -0.332     0.023
New_Housing_unit_started_YoY           -0.0179      0.021    -0.870    0.385    -0.058     0.022
Net_Trade_in_Goods_YoY                  0.0370      0.024     1.540    0.124    -0.010     0.084
Uni_Michigan_Consumer_Sentiment_YoY    -0.0968      0.027    -3.563    0.000    -0.150    -0.044
Industrial_Production_YoY              -0.3135      0.120    -2.621    0.009    -0.548    -0.079
SPX_Index_Return                       -0.0105      0.098    -0.108    0.914    -0.202     0.181
tsy10_tsy2                              0.0044      0.003     1.259    0.208    -0.002     0.011
Bloomberg_10yr_Muni_Index_YTW           0.2189      0.217     1.007    0.314    -0.207     0.645
Non_Farm_Payroll_YoY                    0.3132      0.177     1.773    0.076    -0.033     0.659
Monthly_supply_of_new_houses_YoY        0.0256      0.015     1.686    0.092    -0.004     0.055
Mexico_Consumer_Price_Index_YoY        -0.0143      0.015    -0.959    0.337    -0.043     0.015
Alcoholic_Beverages_YoY                -0.0441      0.121    -0.366    0.714    -0.280     0.192
==========================================================================================
```

$Confusion\ Matrix: \begin{bmatrix} 94 & 2 \\ 2 & 11 \end{bmatrix}$    Prediction Accuracy: 96.33%    F1 Score: 84.62%

In the trained model the Pseudo R-square of 74.5% and log likelihood ratio of 8.6e-24 indicates the model as a whole is statistically significant and it fits significantly better than a model with no predictors. Also, coefficients show with 1 unit increase the Z-score would increase/decrease by the amount of coefficient. After, running the trained model on the test set, it gave an accuracy of 96.33%, but as the response data is unbalanced, I used F1 score to have a true understanding of the accuracy which decreased the model accuracy to 84.62%.

Since the dataset is not large enough, five-fold cross validation is used to make sure the accuracy is reliable, but the result had a lot of variation that indicates the model is not reliable enough. The variation in the accuracy, as also concluded by Puglia and Tucker (2020), is due to rigidity of Probit model. Since Probit is trying to draw a line in feature space by minimizing cost function, there is a trade-off between goodness of fit in some feature space versus poorness of fit in other areas. Therefore, Probit is not good in non-linear space.

Next, I tried Logit model. Logit is also a probability regression model, the difference from Probit is in the S-shaped curve used to constrain probabilities between 0 and 1, also, the process of calculating probabilities is based on linear combinations in Logit, while Probit uses cumulative normal distribution function. The function (Kovacova and Kliestik, 2017) can be defined as:

$$\text{Logit}(P_1) = \ln\left(\frac{P_1}{1-P_1}\right) = f(x,\beta) = \beta_0 + \beta_1 x_1 + \beta_2 x_2 + \cdots + \beta_n x_n$$

(13)

The result from Logit is stated below (Fig 6).

**Fig. 6** *Logit model summary table, out of sample confusion matrix, prediction accuracy, and F1 score*

```
results_logit.summary():
                           Logit Regression Results
==============================================================================
Dep. Variable:              Recession   No. Observations:                  329
Model:                          Logit   Df Residuals:                      316
Method:                           MLE   Df Model:                           12
Date:                Mon, 29 May 2023   Pseudo R-squ.:                  0.7612
Time:                        14:55:59   Log-Likelihood:                -22.300
converged:                       True   LL-Null:                       -93.366
Covariance Type:            nonrobust   LLR p-value:                 2.221e-24
========================================================================================
                                           coef    std err          z      P>|z|      [0.025      0.975]
----------------------------------------------------------------------------------------
Intercept                                -9.0428      2.594     -3.486      0.000     -14.128      -3.958
UKX_Index_Return                         -0.2616      0.171     -1.530      0.126      -0.597       0.073
New_Housing_unit_started_YoY             -0.0492      0.047     -1.042      0.297      -0.142       0.043
Net_Trade_in_Goods_YoY                    0.0490      0.048      1.026      0.305      -0.045       0.143
Uni_Michigan_Consumer_Sentiment_YoY      -0.1823      0.055     -3.345      0.001      -0.289      -0.075
Industrial_Production_YoY                -0.6443      0.249     -2.591      0.010      -1.132      -0.157
SPX_Index_Return                         -0.0081      0.187     -0.043      0.966      -0.375       0.359
tsy10_tsy2                                0.0108      0.008      1.397      0.162      -0.004       0.026
Bloomberg_10yr_Muni_Index_YTW             0.5879      0.453      1.300      0.194      -0.299       1.478
Non_Farm_Payroll_YoY                      0.6809      0.364      1.868      0.062      -0.033       1.395
Monthly_supply_of_new_houses_YoY          0.0469      0.032      1.469      0.142      -0.016       0.109
Mexico_Consumer_Price_Index_YoY          -0.0271      0.029     -0.938      0.348      -0.084       0.030
Alcoholic_Beverages_YoY                  -0.0705      0.227     -0.311      0.756      -0.515       0.374
========================================================================================
```

$Confusion\ Matrix: \begin{bmatrix} 93 & 3 \\ 3 & 10 \end{bmatrix}$    Prediction Accuracy: 94.5%    F1 Score: 76.92%

Although the model had slightly higher Pseudo R-squared (76.12%), tested model on the test set resulted in a lower accuracy and F1 score of 94.5% and 76.92% respectively. Again, when the five-fold cross validation is run the accuracy varied a lot which is an indication of unreliable model.

It worth to note, there are two possible reasons for variation in cross validation accuracy. First, it may be because the dataset is relatively small. Second, the subsets that are used in the cross validation may be small that is not representative of the whole data to train a robust model. As a result, cross validation with different K values is examined and the result had still large variation. Thus, I conclude the issue is arising from the sample size.

**Elastic Net:** Elastic Net is a regression base model that emerged to have the best of both Lasso and Ridge model. It let the data to decide on the loss function, L1 ratio, that the model should be more like a Lasso or Ridge regression.

As Garcia-Nieto, Garcia-Gonzalo, Paredes-Sanchez (2021) explained in detail Lasso and Ridge models, there is a large difference between the two models. Ridge regression is similar to least square regression that minimizes the coefficients $\beta$ with the aim of lowering residual sum of squares (RSS):

$$RSS = \sum_{i=1}^{n}(y_i - \beta_0 - \sum_{j=1}^{p}\beta_j x_{ij})^2$$

(14)

The loss function would be as follow, where $\lambda$ is the complexity or tunning parameter:

$$L^{RR}(\beta) = RSS + \lambda \sum_{j=1}^{p} \beta_j^2$$

(15)

Worth noting, as $\lambda$ shrinkage penalty gets closer to 0, the Ridge regression would be more like least square model, however, as $\lambda$ grows the coefficients would shrinkage towards zero. Unlike least square, Ridge will produce a set of coefficient estimates for each $\lambda$ value.

There is a disadvantage in Ridge regression that it does not set any of coefficients exactly zero. However, in the Lasso regression, that is alternative to Ridge model, it overcomes this problem. It replaces the $\beta_j^2$ term in Ridge with $|\beta_j|$:

$$L^{RR}(\beta) = RSS + \lambda \sum_{j=1}^{p} |\beta_j|$$

(16)

Although the Ridge would shrike the coefficient estimates, Lasso forces some of coefficients to be exactly zero. Therefore, Lasso also performs as variable selection and easier to interpret. However, Lasso variable selection can be too data dependent and unstable.

Elastic Net is constructed to bring the best of both models together. It has two tuning parameters $(\lambda, \alpha)$ with the aim of minimizing the loss function, as follow:

$$L^{Elastic\ Net}(\beta) = \frac{1}{2n}\sum_{i=1}^{n}(y - \beta_0 - \sum_{j=1}^{p}\beta_j x_{ij})^2 + \lambda(\frac{1-\alpha}{2}\sum_{j=1}^{p}\beta_j^2 + \alpha\sum_{j=1}^{p}|\beta_j|)$$

(17)

Where $\alpha$ is between 0 (making the equation equal to Ridge) and 1 (making the equation equal to Lasso).

According to empirical studies, Elastic Net performs well on the data when the predictors are highly correlated. It doesn't simply eliminate the high collinearity coefficients. After training elastic net with five-fold cross validation and testing on the test set, the chosen parameters, importance variable plot, and performance metrics are stated below (Fig 7).

*Fig. 7 Elastic Net variable importance plot, confusion matrix, accuracy and F1 score*

*model chosen L1 ratio and alpha are 0.01 and 0.29 respectively*

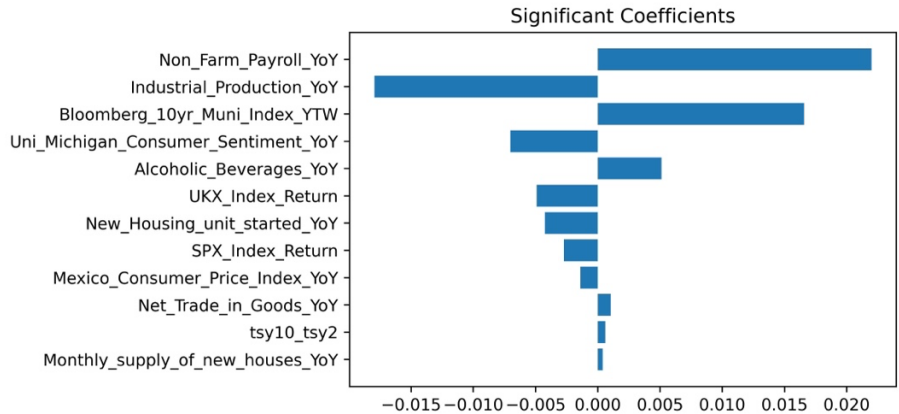

$Confusion\ Matrix: \begin{bmatrix} 96 & 0 \\ 10 & 3 \end{bmatrix}$     Prediction Accuracy: 90.83%     F1 Score: 37.5%

Since the chosen L1 ratio is 0.01, our model is very close to Ridge, also optimization parameter alpha is chosen 0.29 that is where the cross-validation error is minimized. Interestingly, the four most important variables are non-farm payroll, industrial production, 10yr municipal index, and consumer sentiment. Although the tested model accuracy is 90.83%, F1 score is very low 37.5%. By looking at the confusion matrix, the model is predicted all the periods as non-recession except three periods, that is the reason why there is a large difference between accuracy and F1 score; however, it never predicted a non-recession period wrong. Overall, the model did not perform well on the test set as it is predicted mostly non-recession.

**Random Forest:** Random forests (Breiman, 2001) are models famous for dealing with noisy, non-linear, high dimensional problem sets. The proof of their reliability is available in Biau (2012). A Random Forest algorithm creates several decision trees through bootstrap aggregating. Then considering training samples and fit many trees to each sample set of $X = X_1,...,X_N$ with the response variable sample of $Y = y_1,..., y_N$. Then in order to reduces variance and correlation between trees, its samples with replacement and gets the average of trees. When it trains each tree $f_x$ on $X_{1,...,B}$, $Y_{1,...,B}$ (where B is number of trees), a random number of features are considered to make sure all trees are not using most predictive features of the training data during model construction, therefore it will cause to decrease the correlation between the trees. Lastly each tree makes a classification, and the average of trees can be seen as:

$$\mathcal{F}(x) = \frac{1}{B} \sum_{b=1}^{B} f_b(x)$$

(18)

Based on elastic net L1 ratio in the previous method that gave a model closer to ridge, it is understandable that the data is more suffering from high variance and so Random Forest can be useful.

First, Random Forest cross validation with all features is run to find the optimal maximum features that gives lowest mean squared error (MSE). Then, trained the model with max features and evaluated the model (Fig 8).

*Fig. 8* Random Forest selected feature's mean squared error (MSE), variable importance plot, out of sample confusion matrix, accuracy and F1 score

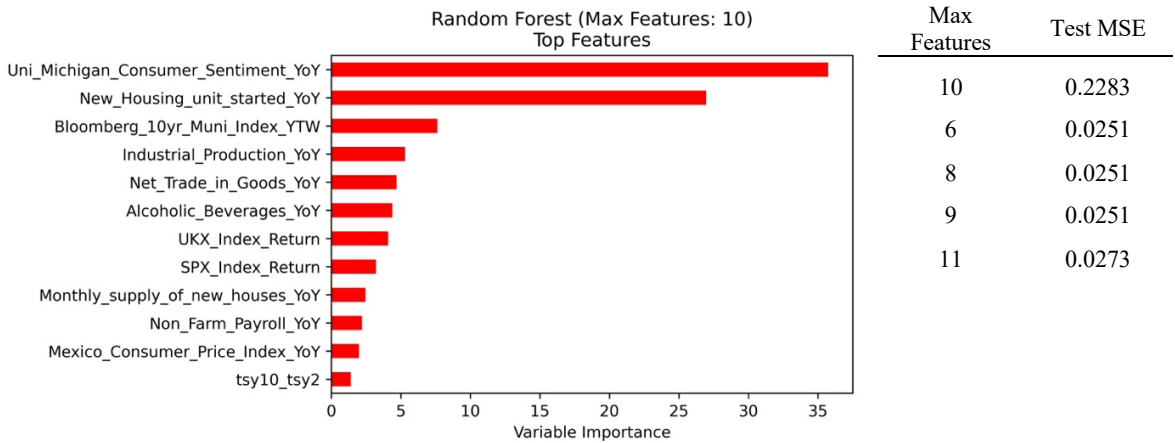

| Max Features | Test MSE |
|---|---|
| 10 | 0.2283 |
| 6 | 0.0251 |
| 8 | 0.0251 |
| 9 | 0.0251 |
| 11 | 0.0273 |

$Confusion\ Matrix: \begin{bmatrix} 95 & 1 \\ 7 & 6 \end{bmatrix}$    Prediction Accuracy: 92.66%    F1 Score: 60%

Max number of features are chosen 10 that giving the lowest MSE, Also, consumer sentiment and new housing units started considered by far the most important variables in trees. In terms of model evaluation on the test set, the prediction accuracy is pretty high 92.66% with F1 score of 60% that shows a better result compared to other models since it recognized six recession periods correctly out of thirteen. Lastly, due to using cross validation, the evaluation is reliable.

**Gradient Boosting:** Boosting works in a similar way as bagging, except that trees are grown sequentially using previous grown tree's information. A commonly used boosting model is Gradient Boosting, it tries to map instances of x to their response variable y by minimizing the expected value of a given loss function *L (y, F(x))*. Gradient Boosting builds an additive function as follows:

$$F_m(x) = F_{m-1}(x) + \rho_m h_m(x)$$

(19)

$$F_0(x) = \underset{\alpha}{\text{argmin}} \sum_{i=1}^{N} L(y_i, \alpha)$$

(20)

Where $\rho_m$ is the weight of the $m^{th}$ random forest function model, $h_m(x)$. Also, $F_0(x)$ is a constant approximation of model that expected to minimize the loss function as follow:

$$(\rho_m, h_m(x)) = \underset{\rho,h}{\text{argmin}} \sum_{i=1}^{N} L(y_i, F_{m-1}(x_i) + \rho h(x_i))$$

(21)

As illustrated by Bentejac, Csorgo, and Martinex-Munoz (2019), a shrinkage value (commonly known as learning rate) is used to regularize gradient boosting.

$$F_m(x) = F_{m-1}(x) + \nu \rho_m h_m(x)$$

$$\nu = (0, 1.0]$$

(22)

The model is run using cross validation and tuned the parameters to find the optimal values for the impact of each tree on the outcome (learning rate) and the number of trees (number of estimators). The result is stated below (Fig 9).

*Fig. 9 Gradient Boosting selected feature's mean squared error (MSE), variable importance plot, out of sample confusion matrix, accuracy and F1 score*

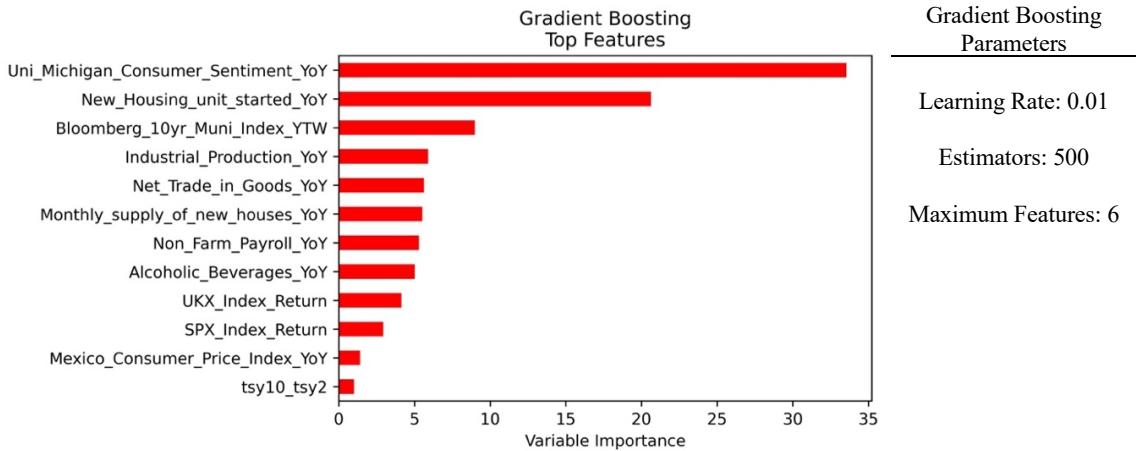

$Confusion\ Matrix: \begin{bmatrix} 96 & 0 \\ 7 & 6 \end{bmatrix}$    Prediction Accuracy: 93.58%    F1 Score: 63.16%

According to five-fold cross validation, 0.01 learning rate, 500 number of estimators with 6 max features are chosen that represented consumer sentiment, new housing unit started, and 10yr municipal index as the most important features in the model. The tested model result was slightly better than random forest (F1 score 63.16%) with the only major difference that the model never predicted a non-recession period wrong. Therefore, when the model predicts a non-recession period, we can be confident about not having a recession, while if the model predicts a recession, around 46% of the times the model is predicting correctly.

**Neural Network:** Since learning information from the previous analysis helped in Gradient Boositng method, I tried Neural Network method that learns complex non-linear functions with the aim to minimize errors, but with the cost of losing interpretability as it trains a complex model. Neural Network consists of layered functions of neurons with connections to each other. The neurons form a complex mechanism that passes signals and learns with some feedback. Each layer can have many neurons that compute a function to pass a signal to next neurons. The connections between neurons have different weights that specify the impact of an input on the next neuron. Neural network set the initial weights all random, but during model training the weights update to minimize the prediction error. By decomposing the Neural Network, we can set the number of neurons, layers, weights, and learning rate of learning mechanism (optimizer) with the aim of minimizing loss in the training process. In this case, the dataset consists of twelve initial variables that I used three hidden layers with each layer converging to eight, five and lastly one outcome; also, the learning rate is set as the previous model on 0.01 with BCEWithLogitsLoss as criterion that computes the loss. The result is stated below.

$$Confusion\ Matrix: \begin{bmatrix} 96 & 0 \\ 14 & 0 \end{bmatrix} \quad Prediction\ Accuracy: 87.27\% \quad F1\ Score: 0\%$$

Although Neural Network is usually used for the accurate prediction, it requires a very large dataset that improves much more by the time. The tested model could not predict any recession periods which caused F1 score of 0%. Ultimately as the dataset has only 438 periods it is not sufficient to train a robust Neural Network model.

## 6. Discussion

In this section I compare the performance of constructed models, and highlighting the results in comparison with the other papers.

There is a need to highlight the fact that the nature of the data is imbalance. There are only 40 recession periods out of 438 total periods (9.1% recession). When failure rates are low specially in an important topic like recession, false discovery rate becomes important. For example, if the model predicts a normal state, but the reality turns out to be a recession, the consequences of false discovery can be massive.

Initially, seven different subset of factors (75 characters) considered, that after Boruta algorithm and correlation matrix, 12 most important variables left for analysis. Among linear models (Probit, Logit, Elastic Net) most important variables are Non-farm Payroll YoY, Industrial Production YoY, and Municipal Index Return, however, it defers in non-linear models (Random Forest & Gradient Boosting) to University of Michigan Consumer Sentiment YoY and Number of New Houses Started YoY. Worth noting, in non-linear models Industrial Production YoY and Municipal Index Return are still in top five most important characters, although their effect is much less than the other two characters mentioned above, also, among 75 factors most of them do not have predictability power.

Some papers such as Gogas (2014) and Fornani (2010) illustrated for a few cycles ahead treasury term spread and fed fund rate have strong capability in comparison with other macroeconomics factors. They pointed out as the time passes from one to three quarters ahead, the predictability power of yield curve slope increases. They believe the effect of financial tightening and economic conditions would be reflected in the yield curve 6-9months ahead. However, in this paper figured in short-time horizon other factors such as labor market, housing market, and consumer sentiment are much more helpful to predict recessions because these factors show the current state of the economy.

For the better comparison, the out of sample overall accuracy and F1 score for each model are provided below (Table 5):

*Table 5: shows out of sample performance summary table per each model*

|  | Probit | Logit | Elastic Net | Random Forest | Gradient Boosting | Neural Network |
|---|---|---|---|---|---|---|
| Accuracy | 96.33% | 94.50% | 90.83% | 92.66% | 93.58% | 87.27% |
| F1 Score | 84.62% | 76.92% | 37.50% | 60% | 63.16% | 0% |

Probit and Logit model's accuracy and F1 scores are higher than other models, however, after running cross validation their accuracy varied a lot. These model performances are not reliable and robust as the high performance is due to sample selection effect. According to Elastic-Net, another linear model, the high variance of the characters and low failure rate in the dataset resulted in a poor performance (F1 score 37.5%).

On the other hand, non-linear classification models, Random Forest and Gradient Boosting, out of sample F1 scores are 60% and 63.16% respectively. Their flexibility caused a more robust and better performance in comparison with linear models. Whenever Gradient Boosting predicted a normal state always no recession occurred, but when the model predicted a recession happens only half of the times a recession occurred. From policy and risk management standpoint, it is very important that the trained model never predicted a normal state and a recession happened. Lastly, Neural Network doesn't show any predictability power as it could not predict any recession periods correctly due to very large sample requirement for building complex connections between neurons.

Over the recent years, many researchers with different approaches have tried to build a model to first understand and then predict recession states. Many used different subset of factors, but the common findings among all are the fact that low failure rate (Error I) caused the prediction to become so hard; besides, machine learning classification models are performing better in economic recession prediction (Puglia and Tucker, 2020; Gogas, 2014; Coulombe and Marcellino and Stevanovic, 2021).

Malladi (2022) investigated both recession and stock market crash, and he found out predicting economic recession is three times harder than stock market crash. He explored 134 explanatory variables and concluded LSVM model is the best performer among machine learning algorithms in three-month ahead recession forecasting, 91.4% correctly predicted the recession periods. However, it failed to predict all the non-recessionary periods. In comparison to the models created in this paper, the best model (Gradient Boosting) predicted only 46.1% of times recession periods correctly but captured all non-recessionary periods.

Since the consequence of predicting a recession period as non-recession can be severe, I believe the model created in this paper is more reliable for short-time horizons to be used among the standard tools using in United States recession forecasting.

## 7. Conclusion

The issue of forecasting recessions has been always important due to its implications in terms of economic policy and economic stability. Since we are living in a data driven world,
it is very valuable to extract information from past data to understand the economic downturns better. For this reason, I started with a large dataset consist of seventy-five explanatory variables from Jan1986, after backcasting the missing data using ARIMA time-series model, I reduced the high dimensional dataset to twelve most relevant characters by Boruta algorithm and correlation matrix. Afterwards, I evaluated Probit,

Logit, Elastic Net, Random Forest, Gradient Boosting, and Neural Network models (Table 5).

Since the response binary variable (recession) is very unbalanced, F1 score helped to get a true understanding about the accuracy of the models. According to the table 5, probability regression models performed better but after running cross validation the result varied a lot that is the indication of unreliability of the Probit and Logit models. Also, Neural Network could not predict any of the recession periods correctly due to the size of dataset as it consists of only 438 periods. In terms of machine learning models, after running cross validation they captured non-recession periods almost correctly (Random Forest failed only on one period), however, they could only recognize half of recession periods. All in all, I believe the Gradient Boosting trained model is performed the best as whenever it predicted a non-recession period no recession occurred. Worth highlighting, as economic sentiments usually have monthly frequency and recession prediction has a low failure rate, more data points would be a major help to enhance the prediction power.

## 8. Reference

1. Coulombe, P.G. and Marcellino, M. and Stevanovic, D. (2021). Can Machine Learning Catch The COVID-19 Recession? *National Institute Economic Review.* Vol 256. PP 71-109. https://doi.org/10.1017/nie.2021.10.
2. Donoghue, E. (2009). Economic Dip, Decline or Downturn? An Examination of The Definition of Recession. *Student Economic Review.*
3. Gogas, P. and Papadimitriou, T. and Matthaiou, M. and Chrysanthidou, E. (2014). Yield Curve and Recession Forecasting in a Machine Learning Framework. *Comput Econ*. https://doi.org/10.1007/s10614-014-9432-0.
4. Layton, A.P. and Banerji, A. (2003). 'What is a recession?: A reprise'. *Applied Economics* 35:1789-97.
5. Loh, Wei-Yin (2014). Fifty years of classification and regression trees. *International Statistical Review*, 34: 329-370.
6. Malladi, R.K. (2022). Application of Supervised Machine Learning Techniques to Forecast the COVID-19 U.S. Recession and Stock Market Crash. *Comput Econ*. https://doi.org/10.1007/s10614-022-10333-8.
7. Nyman, R. and Ormerod, P. (2016). Predicting Economic Recessions Using Machine Learning Algorithms. *ArXiv Working Paper No.* arXiv:1701.01428.
8. Nyman, R. and Ormerod, P. (2020). Understanding The Great Recession Using Machine Learning Algorithms. *ArXiv Working Paper.*
9. Puglia, M, and Tucker, A. (2020). "Machine Learning, the Treasury Yield Curve and Recession Forecasting," *Finance and Economics Discussion Series 2020-038. Washington: Board of Governors of the Federal Reserve System*, https://doi.org/10.17016/FEDS.2020.038.
10. Raschka, S. (2018). Model Evaluation, Model Selection, and Algorithm Selection in Machine Learning. *ArXiv Working Paper.* arXiv:1811.12808.
11. Williams, A. and Miller, M. (2013). Do Stocks With Dividends Outperform The Market During Recessions? *Journal of Accounting and Finance. Vol 13(1).*

12. Schapire, R. and Freund, Y. (1999). A Short Introduction to Boosting. Jour- nal of Japanese Society for Artificial Intelligence, 14(5):771 – 780.
13. Friedman, J. (2001). Greedy function approximation: a Gradient Boosting machine. The Annals of Statistics, 29(5):1189 – 1232.
14. Bentejac, C. and Csorgo, A. and Martinez-Munoz, G. (2021). A Comparative Analysis of XGBoost. *Artif Intell Rev* 54. https://doi.org/10.1007/s10462-020-09896-5
15. Garcia-Nieto, P. Garcia-Gonzalo, E. Paredes-Sanchez, J. (2021). Prediction of The Critical Temperature of A Superconductor by Using The WOA/MARS, Ridge, Lasso and Elastic-Net Machine Learning Technics. Natural Computing and Applications. https://doi.org/10.1007/s00521-021-06304-z
16. Sapra, S. (2005). A Regression Error Specification Test (RESET) For Generalized Linear Models. Economics Buletin. Vol. 3, No. 1 pp 1-6. http://www.economicsbulletin.com/2005/volume3/EB-04C50033A.pdf
17. Kovacova, M. and Kliestik, T. (2017). Logit and Probit Application For The Prediction of Bankruptcy In Slovak Companies. *Equilibrium. Quarterly Journal of Economics and Economic Policy.* Pp 775-791. Doi: 10.24136/eq.v12i4.40
18. Wang, Z. and Li, K. and Xia, S. and Liu. H. (2021). Economic Recession Prediction Using Deep Neural Network. arXiv:2107.10980
19. Kursa, M. and Rudnicki, W. (2010). Feature Selection With The Boruta Package. *Journal of Statistical Software.* Vol. 36. Issue. 11. Doi: 10.18637/jss.v036.i11
20. Liu, C. and Hoi, S. and Zhao, P. and Sun, J. (2016). Online ARIMA Algorithms For Time Series Prediction. In *Proceedings of the AAAI Conference on Artificial Inteligence.* Vol. 30. No. 1. Doi: https://doi.org/10.1609/aaai.v30i1.10257
21. Shiskin, J. (2976). Employment and Unemployment: The Doughnut or The Hole?. *Monthly Labor Review.* 99(2). pp 3-10. http://www.jstor.org/stable/41840118
22. O'Donoghue, E.M.M.A. (2009). Economic Dip, Decline or Downturn? An Examination of The Definition of Recession. Student Economics Review. 23. Pp 3-12.

23. Layton, A.P. and Banerji, A. (2003). What Is A Recession?: A Reprise. *Applied Economics, 35(16).* Pp 1789-1797. https://doi.org/10.1080/0003684032000152853
24. Estrella, A. and Mishkin, F.S.(1996). The Yield Curve As A Predictor of US Recessions. *Current Issues in Economics and Finance. 2(7)*.
25. Chauvet, M. and Potter, S. (2002). Predicting A Recession: Evidence From The Yield Curve In The Presence of Structural Breaks. *Economics Letters. 77(20)*. Pp 245-253. https://doi.org/10.1016/S0165-1765(02)00128-3
26. Fornari, F. and Lemke, W. (2010). Predicting Recession Probabilities With Financial Variables Over Multiple Horizons. *ECB Working Paper.* No. 1255.
27. Stock, J.H. and Watson, M.W. (2006). Forecasting With Many predictors. *Handbook of Economics Forecasting, 1.* pp 515-554. https://doi.org/10.1016/S1574-0706(05)01010-4
28. Liaw, A. and Wiener, M. (2002). Classification and Regression By Random Forest. *R News, 2(3).*pp 18-22.
29. Breiman, L. (2001). Random Forests. *Machine Learning. 45.* Pp 5-32.
30. Biau, G. (2012). Analysis of A Random Forests Model. *The Journal of Machine Learning Research, 13(1).* pp 1063-1095.
31. Ahrens, D. (2007). *Investing In Vice: The Recession-Proof Portfolio of Booze, Bets, Boombs, and Butts.* St. Martin's Press.
32. OEC. (2020). *Observetory of Economic Complexity*. https://oec.world.

## 9. Statements & Declarations

### 9.1 Funding

The authors declare that no funds, grants, or other support were received during the preparation of this manuscript.

### 9.2 Competing Interests

The authors have no relevant financial or non-financial interests to disclose.

### 9.3 Author Contributions

All material preparation, data collection and analysis were performed by Kian Tehranian. Kian Tehranian read and approved the final manuscript.

### 9.3 Data Availability

The dataset used and analyzed during the current study are some publicly available at Federal Reserve Economic Data (FRED), https://fred.stlouisfed.org, the rest is not publicly available due to the fact that Bloomberg Terminal is subscription based, https://www.bloomberg.com/professional/solution/bloomberg-terminal/ . But they are available from the corresponding authors on reasonable request.